\documentclass[12pt,preprint]{aastex}
\usepackage{graphicx}
\usepackage{color}
\usepackage{ulem}

\newcommand {\ea} {{\it et~al.}}
\newcommand {\be} {\begin{equation}}
\newcommand {\ee} {\end{equation}}
\newcommand{\hii}{\ifmmode [\rm{H}\,\textsc{ii}] \else [H~{\sc ii}]\fi}
\newcommand{\Ha}{\ifmmode {\rm H}\alpha \else H$\alpha$\fi}
\newcommand{\Hb}{\ifmmode {\rm H}\beta \else H$\beta$\fi}
\newcommand{\oiii}{\ifmmode [\rm{O}\,\textsc{iii}] \else [O~{\sc iii}]\fi}

\shorttitle{Narrow Line Radio Galaxies}

\shortauthors{Sikora \ea}

\begin{document}

\title{Constraining jet production scenarios by studies of
Narrow-Line-Radio-Galaxies}

\author{Marek~Sikora\altaffilmark{1},
  Gra\.zyna~Stasi\'nska\altaffilmark{2}, 
  Dorota~Kozie{\l}-Wierzbowska\altaffilmark{3},
  Greg~M.~Madejski\altaffilmark{4}, and
  Natalia~V.~Asari\altaffilmark{5, 6}}
  
\altaffiltext{1}{Nicolaus Copernicus Astronomical Center, Bartycka 18,
  00-716 Warsaw, Poland; \tt{sikora@camk.edu.pl}}
\altaffiltext{2}{LUTH, Observatoire de Paris, CNRS, Universit\'e Paris
Diderot, Place Jules Janssen 92190 Meudon, France}
\altaffiltext{3}{Astronomical Observatory, Jagiellonian University,
  ul. Orla 171, 30-244 Krak\'ow, Poland} 
\altaffiltext{4}{Kavli Institute for Particle Astrophysics and
  Cosmology, Stanford University, Stanford, CA 94305, USA}
\altaffiltext{5}{Institute of Astronomy, University of Cambridge, Madingley Road,
Cambridge, CB3 0HA, United Kingdom}
\altaffiltext{6}{CAPES Foundation, Ministry of Education of Brazil, Caixa Postal 250,
Brasilia - DF, 70040-020, Brazil}

\begin{abstract}

We study a large sample of narrow-line radio galaxies (NLRGs) 
with extended radio structures.  Using 1.4 GHz radio luminosities, 
$L_{1.4}$, narrow optical emission line luminosities, $L_{\oiii}$ and $L_{H_{\alpha}}$, as well as 
black hole masses $M_{BH}$ derived from stellar velocity dispersions 
measured from the optical spectra obtained with the Sloan Digital 
Sky Survey, we find that:  
(i) NLRGs cover about 4 decades of the Eddington ratio,
$\lambda  \equiv L_{bol}/L_{Edd} \propto L_{line}/M_{BH}$;
(ii) $L_{1.4}/M_{BH}$ strongly correlates with $\lambda$;
(iii) radio-loudness, ${\cal R} \equiv L_{1.4}/L_{line}$, strongly
anti-correlates with $\lambda$.
A very broad range of the Eddington ratio indicates that 
the parent population of NLRGs includes both radio-loud quasars (RLQs) and 
broad-line radio galaxies (BLRGs). The correlations they obey and 
their high jet production efficiencies favor a jet production model which  
involves the so-called 'magnetically choked' accretion scenario. 
In this model, production of the jet is dominated by 
the Blandford-Znajek mechanism, and the magnetic fields in the vicinity
of the central black hole are confined by the ram pressure of the accretion flow. 
Since large net magnetic flux accumulated in central regions of
the accretion flow required by the model can take place 
only via geometrically thick accretion, we speculate 
that the massive, 'cold' accretion events associated
with luminous emission-line AGN can be accompanied by 
an efficient jet production only if preceded by a hot, 
very sub-Eddington accretion  phase.  

\end{abstract}


\section{INTRODUCTION}

It became clear already shortly after the discovery of first radio galaxies that their
strong radio emission is associated with a presence of luminous optical emission 
lines (Bade \& Minkowski 1954;  Osterbrock 1977; Grandi \& Osterbrock 1978).
Such an association was later confirmed by finding a correlation of radio
luminosities with narrow line luminosities in Fanaroff-Riley type II (FRII, Fanaroff \& Riley 1974) radio sources 
(Baum \& Heckman 1989; Saunders et al. 1989; Rawlings et al. 1989;
Rawlings \& Saunders 1991; Zirbel \& Baum 1995; Willott et al. 1999; 
Buttiglione et al. 2010; Kozie{\l}-Wierzbowska \& Stasi\'nska 2011: KS11).  
Using the narrow-line luminosity as a proxy for the cold accretion disk 
luminosity and the radio luminosity as a proxy of the jet power,  Rawlings 
\& Saunders (1991) found an approximate proportionality between these 
two quantities, with jet powers approaching and in some cases, even 
exceeding the bolometric luminosities of the accretion disks.  

Studies of the correlation of the jet powers with properties of central 
engines using radio and optical luminosities became more thorough and robust 
once methods of the black hole mass estimations have been developed (see, e.g.,  Woo \& Urry 
2002 and refs. therein). This allowed a determination of the properties of 
radio-loud AGN as a function of their Eddington ratio, $\lambda$, defined as the ratio of 
the accretion bolometric luminosity to the Eddington luminosity.  In particular, 
Kozie{\l}-Wierzbowska \& Stasi\'nska (2011) using the black hole mass estimates 
from the stellar velocity disperssion ($\sigma_{*}$) -- BH mass ($M_{BH}$) 
relation (Tremaine et al. 2002) found that 
FRII radio galaxies span about 4 decades of the Eddington ratio and that, when scaled 
by their black hole masses, their radio luminosities correlate with Eddington ratio, in similarity  
to the correlation of absolute radio luminosities vs. absolute narrow emission 
line luminosities.  One might consider this similarity as resulting from 
the fact that distribution of the black hole masses of  FR II sources is rather narrow, 
with  majority of them  within the range $10^8-10^9 M_{\odot}$. However,
since the  physics of the jet production is linked to the 
Eddington-scaled accretion rate rather than to its absolute value and  
powers of jets produced with the same efficiency scale with the black hole mass, the 
'primary correlation' to be considered should be the one between 
the Eddington-scaled luminosities.  

Such a correlation, however, contradicts the predictions of jet models which 
relate the strength of the central poloidal magnetic fields (those threading 
innermost portions of accretion disk and the black hole) with maximal pressure in the 
disk. According to the standard accretion disk theory, innermost portions 
of disks accreting at a rate corresponding to the Eddington ratio in the 
range $10^{-4} - 1$ are radiation pressure dominated.  Since this pressure does 
not depend on the accretion rate, the powers of jets - believed to be initially dominated 
by Poynting flux - are not expected to depend on the accretion rate either. In consequence,
such models predict maximal jet powers $\sim 100$ times smaller than those 
observed in RLQs with extended radio structures 
(see, e.g., Ghosh \& Abramowicz 1997).  

The model which {\sl can} account for the energetics of the most powerful jets
and explain the observed radio - optical correlation is the one based on
the so-called magnetically arrested/choked accretion flows (Narayan et al. 2003;
Igumenshchev 2008; McKinney et al. 2012).  In such a model, the amount of 
the net magnetic flux amassed in the central region is so large that 
innermost portions of accretion disks are dynamically affected by central 
magnetic fields.  In this regime, the accretion onto a black hole proceeds via 
interchange instabilities ($\equiv$ magnetic Rayleigh-Taylor instability: 
Stone \& Gardiner 2007 and refs. therein) and magnetic flux threading the black
hole is supported by the ram pressure of the accreting plasma.  

In order to directly confront this model with observations, 
we expand the sample of radio galaxies studied by KS11 by 
including other types of radio morphologies, investigate the correlation of
radio-loudness  ${\cal R}$ (radio-to-optical luminosity ratio) vs. $\lambda$, and 
investigate the dependence of source sizes on the Eddington ratio. The paper is 
organized as follows:  in Section 2 we describe our sample selection and data 
reduction and analysis;  in Section 3 we present results of our analysis of optical and radio 
based correlations;  in Section 4 we investigate a consistency of these results 
with the jet production model which involves 'magnetically-chocked' accretion 
scenario.  Our main conclusions are listed in Section 5.

Throughout the paper we assume a $\Lambda$CDM cosmology with 
$H_0 = 71 \ {\rm km} \ {\rm s}^{-1} \ {\rm Mpc}^{-1}$, $\Omega_m=0.27$,
and $\Omega_{\Lambda}=0.73$.  
 
\section{THE DATA}

\subsection{The sample}

To select our sample, we proceeded in a manner analogous to that described in KS11.
Since we are primarily interested in radio galaxies with elongated structures, we 
started with Cambridge radio catalogs accessible by Vizier where such objects are well 
defined: 3C (Edge et al. 1959; Bennett 1962), 
4C (Pilkington \& Scott 1965; Gower et al. 1967), 
5C (Pearson 1975; Pearson \& Kus 1978; Benn et al. 1982; 
Benn \& Kenderdine 1991; Benn 1995), 
6C (Baldwin et al. 1985; 
Hales et al. 1988, 1990, 1991, 1993ab), 
7C (Hales et al 2007), 8C (Rees 1990; Hales et al. 1995) and 
9C (Waldram et al. 2003). 
We considered all the radio sources from  these catalogs and cross-identified 
them in an automatic fashion with the sample of 926246 galaxies with optical 
spectra from the Sloan Digital Sky Survey (SDSS) DR7 main galaxy sample 
(Abazajian et al. 2009).  
Taking into account the sometimes large positional uncertainties in 
the Cambridge radio catalogs, we adopted a maximum distance 
between the position given for the radio source and that given for the optical galaxy varying from 0.2 arcmin to 1 arcmin, depending on 
the catalog. We thus obtained a list of 2633 radio galaxies with available 
SDSS spectra. The final identification was done using the NVSS 
(Condon et al. 1998) and FIRST (Becker et al. 1995) radio maps, which 
have much better spatial resolution than the Cambridge catalogs, 
and have been obtained at the same frequency of 1.4 GHz. For all the pre-selected 
objects we constructed jpg images superimposing NVSS or FIRST contours on SDSS 
images in the $r$ band. The images are centered on the galaxy which is supposed to 
correspond to the radio source. By visual inspection of all these images, 
we found that 307 cases were actually misidentifications (among which 
many of them corresponding indeed to a galaxy but for which the 
SDSS spectrum was that of another, nearby galaxy), and 14 corresponded 
to spiral galaxies whose radio emission is produced over the entire 
disk, most likely by star formation.  For the remaining 2042 radio sources we carried 
out a morphological classification by eye resulting in the following subclasses:
\begin{itemize}
  \item FR I type radio sources, considered as those where the maximum 
    brightness of the lobes is closer to the center than to the extremity,
  \item FR II type, where the maximum brightness of the lobes is closer to the extremity,
  \item FR I/II type, where one lobe is of FR I type and the other of FR II type,
  \item double-double radio sources where two pairs of coaxial lobes are detected,
  \item X-shape radio sources with two pairs of lobes forming an X-shape structure,
  \item one-sided radio sources showing only one lobe,
  \item ``elongated'', i.e. radio sources that do not appear point-like, but 
    whose angular size does not allow us to classify them more accurately, 
   \item radio sources, for which no morphological class could be assigned because of an atypical, 
     irregular shape
\item compact, unresolved radio sources.
\end{itemize}
This morphological classification, which is purely subjective, was carried out 
independently by two of the co-authors, DKW and GS. For about 15\% of the objects 
we considered, our initial classifications diverged, although we agreed on 
our final classification. 
For further considerations in this paper, we restricted our sample to objects 
that clearly show the presence of radio structures associated with radio lobes
and/or jets, i.e. FRII, FRI, FRII/FRI, 
double-double, X-shape and one-sided. 

We excluded the radio sources 
whose parent galaxies have a redshift larger than 0.4, in order to insure that 
the H$\alpha$ emission line -- which is crucial for our study -- falls within 
the SDSS spectral range.  
We also excluded those radio sources where a broad component 
was clearly seen in the hydrogen emission lines. This allows us to make more accurate computations
of narrow line luminosities as it avoids issues related to the decomposition 
of the narrow and the broad components. In addition, this limits our sample to objects which according to the Unified Scheme 
are observed at large inclination angles and by this step, we minimize viewing 
angle biases in our sample, in particular regarding source sizes.
Finally, objects corresponding to galaxies where no emission lines were detected (after 
processing with the STARLIGHT code, Cid Fernandes et al. 2005, see below)  were obviously removed 
from the sample, since all the considerations in this paper make use of emission line fluxes. 
After all these cuts, the entire sample on which this paper is based consists of 404 objects. They are listed in Table 1 (accessible only on-line), together with the properties that will be used.

There are some differences between our starting sample and the catalog of elongated radio sources published by Lin et al. (2010). The latter was assembled via cross correlation of the SDSS DR6 with NVSS and FIRST.  Those radio surveys have the advantage of being more homogeneous and deeper than the Cambridge catalogs. Indeed, the limiting radio flux density at 1.4 GHz in Lin's et al. sample is 3 mJy while that of the 3C catalog  (when rescaled to 1.4 GHz) is 2 Jy, and those of the 4C-9C catalogs range between 400 mJy and 20 mJy (except for the 5C catalog which reaches 1.5 mJy but in very limited zones of the sky). One may wonder how many objects we are missing by using the Cambridge catalogs as a starting point rather than the NVSS one. It turns out that the number should not be very large, since our sample is limited to a redshift of 0.4, contains only objects with FRI, FRII and related morphologies which are the most luminous radio sources and, in addition, includes only objects with emission lines. All our objects have 1.4 GHz luminosities larger than $10^{24}$ W~Hz$^{-1}$. In Lin's et al. catalog, there are only 120 objects with radio luminosities smaller than that out of a total of 1040 objects. This implies that, in our sample, we miss at the very most 12\% of objects on the low luminosity side. 
On the other hand, for some reason, our starting sample contains 112 objects that fulfil all the selection criteria of  Lin et al.  but do not appear in their catalog (such as, e.g. the well-known objects 3C198, 4C+00.56 which are  bright in radio and have large angular sizes). 
Another difference between our approach and the one of Lin et al. is that, while they attempted to define an objective way to trace the galaxy population smoothly from FRI sources to FRII, (as opposed to a sharp and perhaps arbitrary distinction between type I versus type II), we adhered to  more a classical morphological classification paying attention to other types than just FRI and FRII.

\subsection{Data processing}

The 1.4 GHz radio luminosities, 
$L_{1.4}$, were obtained for each source from the sum of 1.4 GHz fluxes of each of its components listed 
in the NVSS catalog, including the central, compact point source.  The NVSS catalog was used here in order to avoid a 
flux loss from the extended and faint components missed in the FIRST catalog.  

The determination of the angular sizes of the radio sources depended on the morphologies. 
For FRII radio sources,  the angular sizes were defined as the distances between 
the hot spots or the most distant bright structures in opposite lobes. They were 
estimated from the FIRST maps, if available, or from the NVSS maps otherwise.  
The determination of the sizes of FRI radio sources is less straightforward, since 
lobes and plumes of the FRI radio galaxies fade away with the distance from the radio core.
For each FRI source in our sample, we obtained
the 3 rms contour using the FIRST map and we 
took for the source size the largest extent of this contour, measured along a straight line 
crossing the radio core (if present). This procedure
works well with straight sources. In the case of bent  sources, the size 
determined in that manner is an underestimate. For one-sided sources, we adopted 
a procedure similar to the one for FRI sources to determine the angular size 
of the only lobe. The linear (projected) sizes were then determined from 
the observed angular sizes using the redshift of the corresponding galaxy 
as obtained from the SDSS. To meaningfully compare the  sizes of all 
the sources, we divided by two the sizes of all the two-sided radio-sources: 
the result is called the characteristic lobe-size in the remaining of this work.

The line luminosities of the galaxies associated with the radio sources 
are fundamental for our work. It is therefore important to determine them 
in the best possible way. For a removal of the stellar features from the 
observed optical spectrum, the best approach is to fit the observed continuum -- excluding 
the spectral zones where emission lines are expected -- with a composite 
stellar population obtained by spectral synthesis and subtract it from the 
entire observed spectrum. What remains is the pure emission line spectrum, 
whose intensities can then be measured.   
As in KS11, we have taken the 
 \oiii 5007, H$\alpha$ and H$\beta$ line fluxes  from 
the STARLIGHT database\footnote{see http://www.starlight.ufsc.br}, 
where they have been obtained precisely in this manner.

As argued in KS11,
we consider $L_{\Ha}$ to be a much better measure of the bolometric luminosity 
of the AGN than the commonly used $L_{\oiii}$, because it does not depend on the
ionization state. This has also been shown by Netzer (2009).
The plots presented below are, however, presented in pairs, 
with one plot using $L_{\Ha}$ and the other using $L_{\oiii}$, for easy 
comparison with works by other authors.

In only about 20\% of objects in our sample are both the H$\alpha$ 
and H$\beta$ line fluxes measured with sufficient accuracy to allow 
a meaningful estimation of the extinction from their ratios.  In the 
majority of those cases, the extinction $A_{\rm V}$ is smaller than 1 
although in a couple of cases it reaches values of up to 4. Note that 
we find no correlation between $A_{\rm V}$ and the radio luminosity 
and the luminosity in the lines. There is then no other way than to 
ignore extinction in our work, if we want to work with a sample with 
significant size. Ignoring extinction will then simply add some dispersion 
to the properties derived from the comparison of optical and radio data, 
which is not really an issue in our work. 

The black hole masses of the galaxies were estimated from the observed 
stellar velocity dispersion given by the SDSS, $\sigma_{*}$,  using the 
relation  by Tremaine et al. (2002): 
\begin{equation}
\label{eq: MBH}
{\rm log} M_{\rm BH} = 8.13 + 4.02\, {\rm log} (\sigma_{*} / 200\,{\rm km\,s^{-1}}).
\end{equation}
In considerations involving black hole masses we disregard cases with  $\sigma_{*} < 60$\,km\,s$^{-1}$ as well as cases 
where the signal-to-noise ratio of the SDSS spectrum at 4000 \AA\ is smaller 
than 10, to ensure that the estimate of $M_{\rm BH}$ is not significantly 
affected by observational errors. 

\section{RESULTS}

\subsection{Radio vs. optical luminosities}

\begin{figure}
\centering
\includegraphics[width=0.95\textwidth]{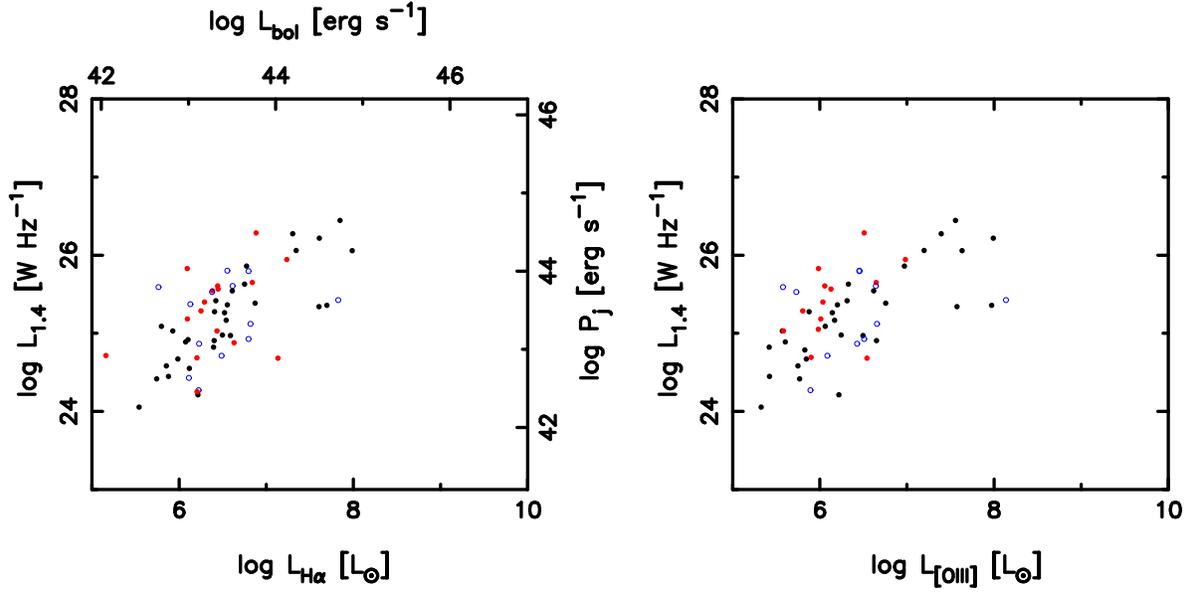}
\caption{The radio luminosities 
 $L_{1.4}$ as a function of the optical line luminosities $L_{\Ha}$ (left panel) and $L_{\oiii}$ (right panel) for the subsample within the redshift range $0.05 < z <0.1$.  FRI types are represented by filled red circles (grey in the printed edition), FRII types by filled black circles, 
and the remaining types, i.e. FRI/II, double-double, X-shape and one-sided in open blue circles (grey in the printed edition). In the left panel, the values of the bolometric luminosity and of the radio luminosity $P_j$ in ergs s$^{-1}$
calculated for $f=3$ are also indicated. }
\label{fig:fig1a}
\end{figure}

\begin{figure}
\centering
\includegraphics[width=0.95\textwidth]{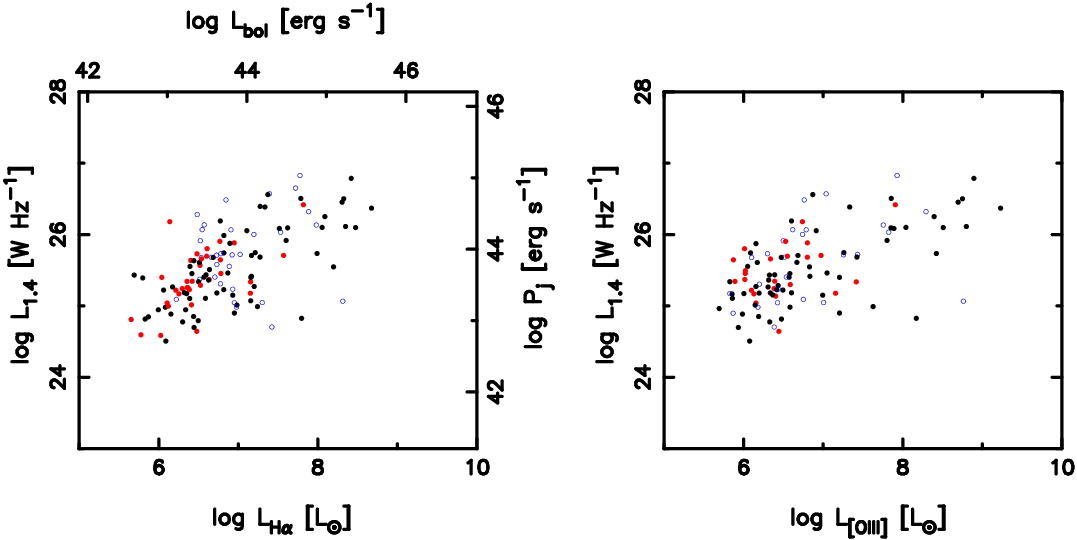}
\caption{Same as Fig. \ref{fig:fig1a} for the redshift range $0.1 < z <0.2$. }
\label{fig:fig1b}
\end{figure}

To  compare radio and optical luminosities, we consider three narrow redshift bins, $0.05 < z < 0.1$, $0.1 < z < 0.2$ and $0.2 < z < 0.4$, to reduce the effect of the common parameter -- distance -- between these two quantities.
Figures \ref{fig:fig1a}, \ref{fig:fig1b} and \ref{fig:fig1c},  present the three subsamples in 
the $L_{1.4}$ - $L_{\Ha}$ and $L_{1.4}$ - $L_{\oiii}$ planes.  Radio galaxies 
of FRI type are represented by filled red (grey in the printed edition) circles, FRII types by filled black circles, 
and the remaining types, i.e. FRI/II, double-double, X-shape and one-sided in open blue (grey in the printed edition) circles. 
As it is apparent from these figures, the radio luminosities clearly correlate with the
narrow-line luminosities. The Spearman rank correlation coefficients are $r_S = 0.62$ for {\Ha} 
lines and $r_S = 0.58$ for {\oiii} lines in Fig. \ref{fig:fig1a}, $r_S = 0.57$ and $r_S = 0.55$ in Fig. \ref{fig:fig1b} and $r_S = 0.50$ and $r_S = 0.46$ in Fig. \ref{fig:fig1c}
\footnote{Rather than the widely used Pearson 
correlation coefficient to measure the strenghts of correlations, we prefer to consider 
the Spearman rank correlation coefficient which is appropriate for all the diagrams 
presented in this study. Indeed, the use of the Pearson correlation coefficient may 
lead to spurious interpretations of correlations in the case where both  variables 
are scaled by a common factor (see e.g. Dunlap et al 1997, Barraclough 2007). }. 
In  the left panels of Figures \ref{fig:fig1a}, \ref{fig:fig1b} and \ref{fig:fig1c},  we also labelled the axes in units of  $P_j$ 
and $L_{bol}$, where 
$P_j$ is the jet power  and $L_{bol}$ is the AGN bolometric
luminosity. They are obtained by using the following 
conversion formulae:

\be L_{bol} = 2.0 \times 10^3 L_{H\alpha}  = 
7.8 \times 10^{36} L_{H\alpha}[L_{\odot}] {\rm erg~s^{-1}} 
\, , \label{Lbol} \ee
(Netzer 2009), and
\be P_j = 1.6 \times 10^{18} (f/3)^{3/2} L_{1.4}[{\rm W~Hz^{-1}}] {\rm erg~s^{-1}}
\, , \label{Pjet} \ee
the latter being taken from Willott et al. (1999) with the following modifications: (i) conversion from 151 MHz to 1.4 GHz assuming  
the spectral index $\alpha=0.8$
($L_{\nu} \propto {\nu}^{-\alpha}$); (ii) replacement of $P_j \propto L_{\nu}^{6/7}$ by
 $P_j \propto L_{\nu}$, the latter taken with a  normalization factor giving
equality of both at  $L_{1.4} = 10^{26} {\rm W~Hz^{-1}}$. 
With this  normalization the modified formula 
leads to overestimation of a jet  power by a factor
$1.4$ for $L_{1.4} = 10^{27} {\rm W~Hz^{-1}}$ and underestimation by a similar 
factor for $L_{1.4} = 10^{25}{\rm W~Hz^{-1}}$. These differences are not 
substantial when compared with uncertainties of an original formula 
expressed via the parameter $f$, with its $1-20$ range. The figure was made adopting $f$=3. 
One can see that even for low values of $f$  the jet powers of many objects 
exceed their bolometric luminosities. 

\subsection{Radio luminosities vs. Eddington ratio}

In Figure \ref{fig:fig2} we plot the radio vs. emission line luminosities 
normalized by the black hole mass. 
The correlation is strong, with a Spearman rank correlation coefficient 
$r_S = 0.77$ and $r_S = 0.71$ for  \Ha\ and  \oiii\ lines, respectively. 
Noting that $L_{line}/M_{BH}$ provides the proxy for the Eddington ratio
defined to be $\lambda \equiv L_{bol}/L_{Edd}$ and using the conversion 
formulae (\ref{Lbol}) and (\ref{Pjet}), one can see in the left panel of Figure \ref{fig:fig2} that 
NLRG cover about 4 decades of $\lambda$,
from $\lambda \sim 10^{-4}$ up to $\lambda=1$. The figure shows that our sample is dominated by objects
with $\lambda$ spanning the range of $10^{-4}-10^{-2}$. These AGN 
are optically too weak to be considered as hidden quasars, but nonetheless,  
being emitters of strong and high excitation lines are expected to be
powered like  quasars  by cold accretion disks and, if not obscured by torus, 
would appear to us as BLRGs (Barthel 1989; Urry \& Padovani 1995). However, one should note that the division 
of AGN for RLQs and BLRGs does not have any physical grounds,
they form continuous 'Eddington ratio sequence' with no signs of an accretion
mode change (Sikora et al. 2007).\footnote{Historically the AGN  division to RLQs and BLRGs was
related to the stellar vs. fuzzy optical appearance of host galaxies, presently 
it is usually related to the specific  value of the AGN absolute 
optical magnitude, e.g. $M_B = -23.0$ as in the V\'eron-Cetty \& V\'eron 
catalogs (1993).} 

The deficiency of AGN with $\lambda> 0.01$ in our sample confirms the 
earlier indications of rarity of very high accretion rate sources at low redshifts
located in massive galaxies (see, e.g., Kauffmann et al. 2008), while the presence of 
several FRI RGs at $\lambda>0.01$ is consistent with a direct finding 
by Heywood et al. (2007) that radio morphologies of type FRI -- which are usually 
associated with low luminosity RGs -- do happen in quasars as well.

\subsection{Radio-loudness vs. Eddington ratio}

In Figure \ref{fig:fig3} we plot the dependence of $L_{1.4}/L_{line}$ on $L_{line}/M_{BH}$,
where $L_{1.4}/L_{line}$ can be  considered to be the  proxy of the radio-loudness
defined as the radio-to-optical flux ratio.
Our results show a significant negative correlation of radio-loudness with 
the Eddington ratio, with  a Spearman rank correlation coefficient $r_S = -0.54$
and $r_S =-0.63$ when using the \Ha\  and \oiii\ lines, respectively. Such an anti-correlation 
was discovered previously by Ho (2002) for radio quiet AGN  and by Sikora et al. (2007)
for radio-loud AGN, however, the statistics of their studies was too poor to
claim  its presence in the sample when limited only to strong-emission-line objects.

In the left panel of Figure \ref{fig:fig3} we have also indicated the values 
of $P_j/L_{bol}$ and $\lambda$ on the axes. We can see that most objects with $\lambda < 0.01$
have jet powers exceeding bolometric luminosity of their AGN,
those  with lowest $\lambda$'s  even by a factor larger than $10\, (f/3)^{3/2}$ 
(see Eq. \ref{Pjet}). 

\subsection{Sizes}

Figure \ref{fig:fig4}  shows the sizes of the radio lobes as a function of 
the Eddington ratio obtained using \Ha\ (left panel) and \oiii\ (right 
panel). To our knowledge, this is the first time that such a diagram 
is shown. There is only a weak correlation, with  a Spearman rank 
correlation coefficient $r_S = 0.34$ and  $r_S =0.27$ when using 
the \Ha\  and \oiii\ lines, respectively.
This suggests that the expansion of radio sources 
is not accompanied by monotonic, long term changes of the accretion rate, 
and that product of expansion velocity multiplied by the source life-time does 
not depend on the Eddington ratio. Furthermore, the fact that the 
source sizes show no dependence with morphological type suggests that 
radio morphologies do not form any evolutionary sequence.
However one cannot exclude ``switches'' between different morphologies 
caused by  modulations of the jet power and the jet direction.  

\section{DISCUSSION}

Amongst all AGN, the most spectacular from the observational standpoint, yet most challenging 
theoretically -- are probably those associated with extended, luminous 
radio structures.  They appear to us as BLRGs and RLQs if oriented with 
respect to our line of sight such that their nuclei are not obscured by 
dusty tori, and as NLRG otherwise. Their radio and optical luminosities  
imply an efficient energy transport from nuclei to radio-lobes via
narrow relativistic jets at a rate very often exceeding 
the bolometric luminosities of their host nuclei (Rawlings \& Saunders 1991;
Ghisellini et al. 2010; Fernandes et al. 2011; 
Punsly 2011; and Section 3 in this paper).  

As it was demonstrated by Tchekhovskoy et al. (2011) and McKinney et 
al. (2012), such powerful jets can be produced in the scenario which 
involves magnetically arrested/choked accretion flows. Such an accretion 
mode can well take place in the innermost portions of a disk, when the 
amount of the net magnetic flux, $\Phi$, amassed in the central region 
is larger than the maximal flux which can be imparted on that region 
around the black hole by the ram pressure of accreting plasma, 
\be \Phi_{BH} = \phi_{BH} R_g (c \dot M)^{1/2} \, , \label{PhiBH} \ee
where $\dot M$ is the accretion rate, $R_g= GM_{BH}/c^2$, and $\phi_{BH}$ is 
the dimensionless factor called by Tchekhovskoy et al. (2011) the 
``dimensionless magnetic flux.''  The value of $\phi_{BH}$ depends on the 
details of the model and according to the numerical simulations by McKinney 
et al. (2012) is typically on the order of $50$. 
\footnote{Note that dimensionless magnetic flux defined by 
McKinney et al. (2012) and denoted by $\Upsilon_{BH}$
is lower by a factor 5, and is found to be typically of the order of $10$.} 
Jets powered by rotating black holes threaded by such magnetic flux appear to 
gain powers (Blandford \& Znajek 1977; 
Tchekhovskoy et al. 2010) 
\be P_{j}  \simeq  4.0 \times 10^{-3} \, {1 \over c} \, 
\Phi_{BH}^2 \Omega_{BH}^2 f(\Omega_{BH}) \sim
10 \, (\phi_{BH}/50)^2 \, x_a^2 f(x_a) \dot M c^2  \, , \label{PBZ} \ee
where 
\be x_a \equiv R_g \Omega_{BH}/c =  [2(1+\sqrt{1-a^2})]^{-1} \, a \ee
\be f(x_a) \simeq 1+ 1.4 x_a^2 - 9.2 x_a^4 \ee
and  $a \equiv  J_{BH}/J_{BH,max} = c J_{BH}/GM_{BH}^2$ is the
dimensionless angular momentum of a BH, commonly named 'spin'.
For maximal black hole spins, $a \sim 1$ ($\to x_a \sim 1/2$), 
\be P_{j,max} \simeq 1.9 (\phi_{BH}/50)^2 \dot M c^2 =
 19 (\phi_{BH}/50)^2 L_{bol} / (\epsilon/0.1) \, . \label{PBZmax} \ee 
where $\epsilon=L_{bol}/(\dot M c^2)$ is the radiation efficiency of 
an accretion disk.  The jet power may also contain a contribution 
from the accretion flow. However, 
as it  was shown by McKinney et al. (2012) this contribution is never dominant 
and therefore will be ignored in our further discussion.  
As it can be verified using Eqs. (\ref{PBZ}) and (\ref{PBZmax}) and 
Figures \ref{fig:fig1a}, \ref{fig:fig1b}, \ref{fig:fig1c} and \ref{fig:fig2}, the model predicts domination of jet powers 
over AGN bolometric luminosities even for moderate spins.  

As the next step, we will investigate whether -- and how -- such a model 
can explain the anti-correlation of radio-loudness with Eddington ratio shown 
in Fig. 5.  We have
\be {\cal R} = L_{1.4}/L_{line} \propto P_j/L_{bol} = P_j/(\epsilon 
\dot M c^2) = \eta_j/\epsilon \, \ee
where $\eta_j \equiv P_j/\dot M c^2$ is often referred to as the jet production 
efficiency. Hence the negative correlation of ${\cal R}$ with $\lambda$ 
may potentially arise from the respective dependencies of $\eta_j$ and/or $\epsilon$ on
$\lambda$.  According to Eq. (\ref{PBZ}), the dependence of $\eta_j$
on $\lambda$ can eventually result from the dependence of the spin on $\lambda$.
However that would require a negative correlation of spins with $\lambda$,
which is the opposite of what one might expect by noting that the black holes are span up more
efficiently for larger rather than smaller accretion rates. 
Hence we are left  with the option that the disk accretion efficiency decreases
with the decrease of the Eddington ratio, i.e. that $\epsilon$ correlates with
$\lambda$. 

The correlation  can be explained if assuming that all 
cold accretion episodes start with a similar total net magnetic flux $\Phi$, 
which is sufficiently large to exceed $\Phi_{BH}$ for any accretion rate 
(see Eq. \ref{PhiBH}). In such a case, the dynamical dominance of magnetic field 
over the accreting plasma extends up to $R_{m} \gg R_{in}$,
where $R_{in}$ is the inner edge of a standard accretion disk not affected 
by magnetic fields, and  $R_{m}$, sometimes referred to as the magnetospheric
radius, is the size of the region within which the magnetic flux $\Phi$ is 
enclosed. Since below that radius the accretion proceeds via interchange 
instabilities, no significant optical-UV radiation is expected to be produced
within this region. Hence, being 'truncated' at $R_m$, the accretion disk is 
expected to have radiation efficiency $\epsilon \sim R_g/R_{m}$.  
Noting that for a given black hole mass $R_m/R_g \propto (\Phi)^{4/3} {\dot m}^{-2/3}$ 
(Narayan et al. 2003) where $\dot m \equiv \dot M c^2/L_{EDD} = \lambda/\epsilon$, 
one can find that $\epsilon \propto \lambda^{2/5}$ and therefore that   
\be {\cal R} \propto \eta_j/\epsilon \propto \lambda^{-2/5} \, .\ee
Some dispersion is expected to be imposed on this relation 
by distributions of black hole masses and magnetic fluxes (unless 
$\Phi \propto M_{BH}^{3/2}$).  

The  large ratios $R_m/R_g$ for AGN with low values of $\lambda$ are 
consistent with detailed observations of some individual BLRGs which 
strongly suggest existence of the truncation radius in their accretion disks  
(Eracleous et al. 2000; Grandi \& Palumbo 2007; 
Sambruna et al. 2009; Tazaki et al. 2010; Cowperthwaite \& Reynolds 2012).
In those papers, the disk truncation radii inferred from observations 
were interpreted as an effect of obscuration of the central region by Thomson
thick corona, or as a transition to the advection dominated  accretion flows.
Since it predicts large truncation radii as determined  by $R_m$, the model can 
explain the cases of objects with $P_j/L_{bol} > 10$ (see Punsly 2011 and 
refs. therein) without 
the necessity to postulate the jet production efficiency significantly greater than 
unity.\footnote{Formally $\eta_j$ can exceed unity, which simply would mean 
that the rate of extraction of the black hole's rotational energy is larger than the 
rate of the energy inflow (Tchekovskoy et al. (2011).  However as most recent
simulations of McKinney et al. (2012) indicate, it is rather difficult
to   achieve such a solution.}

The scenario presented above needs to be verified by demonstrating 
how such large net 
magnetic flux can be assembled in the central regions  of AGN.
This problem was raised by Lubow et al. (1994), who showed that it is possible 
only if the magnetic Prandtl number is $\le H/R$, where $H$ is the height of 
the disk at a distance $R$ from the black hole center. Since the Prandtl number
predicted in fluids with isotropic turbulence is 
expected to be on the order of unity (Parker 1979), the above condition 
is rather difficult to satisfy in cold, geometrically thin disks (Livio 
et al. 1999; Cao 2011).  This problem can be alleviated, if one considers a 
possibility of an accumulation of a large magnetic flux in the central 
regions of AGN at the onset, or even prior to the period of the cold accretion 
phase. Such an accumulation could occur by dragging magnetic fields 
by geometrically thick, advection dominated accretion flows, such as those
predicted for super-Eddington accretion rates (Jaroszy\'nski et al. 1980;
Beloborodov 1998; Abramowicz 2005) and also for very sub-Eddington accretion 
rates 
(Ichimaru 1977; Rees et al. 1982; Narayan \& Yi 1994, Abramowicz et al. 1995). Since our studies of 
source sizes do not indicate any evolutionary trends of expansion of sources 
 with decreasing Eddington ratio (Fig. \ref{fig:fig4}),
the second option seems to be favored. This is also supported by probabilistic
arguments. For a given frequency  of the cold accretion events, which in
turn are very likely to be triggered by mergers of giant ellipticals with 
less massive, cold gas - rich galaxies (see Ramos Almeida et al. 2012 and refs. 
therein), the fraction of those events 
accompanied by production of powerful jets is predicted to correspond
to a reasonably significant fraction ($\sim 3-10$\%)  of giant ellipticals 
to be in 
radio-active states (Sadler et al. 1989; Donoso et al. 2009; 
van Velzen et al. 2012).
These ellipticals are most likely powered by the  Bondi accretion of hot 
interstellar gas (Burns 1990; Hardcastle et al. 2007; Tasse et al. 2008;
Dunn et al. 2010; Werner et al. 2012).  For typical intensities of interstellar magnetic 
fields in such  galaxies, $\sim 10\mu$G, and their coherence scales, 
$\sim 100$ parsecs (Moss \& Shukurov 1996; Mathews \& Brighenti 2003 and refs. 
therein), the accumulation magnetic fluxes in the central regions of such objects  
needed to exceed $\Phi_{BH}$ for any accretion rate 
can occur within a small fraction of the Hubble time. 

Obviously objects with luminous extended radio-sources such as 
investigated by us NLRGs are exceptional and are only a minority among 
the radio-loud AGN
(see, e.g., de Vries et al. 2006). Most of radio-loud AGN are compact or posses 
weak, diffuse, oval, or more complex extended radio structures. 
They usually  have $P_j \ll L_{bol}$ and therefore
do not require accumulation of very large magnetic fluxes and invoking the magnetically choked accretion flows. 
Production of jets in such objects may proceed via  variety 
scenarios, including those where jets are directly  launched by  
accretion disks (see, e.g., Blandford \& Payne 1982).

\section{CONCLUSIONS}

Main results of our studies of radio and spectroscopic optical 
properties of NLRGs at $z < 0.4$ can be summarized as follows:

\noindent
$\bullet$
Radio luminosities are found to correlate strongly with narrow line 
luminosities.  When converted to the jet powers and AGN bolometric luminosities, 
they indicate that the jet kinetic energy often exceeds the radiative 
output of accretion flows;

\noindent
$\bullet$
NLRGs cover about 4 decades of the Eddington ratio, $\lambda$, with most of 
them having $\lambda < 0.01$. This indicates that their parent population is 
dominated by BLRGs. Together with RLQ they form a continuous Eddington ratio
sequence;

\noindent 
$\bullet$
The 'radio-loudness', ${\cal R} = L_{1.4}/L_{line}$,
shows a strong negative correlation with $\lambda$. This would imply
that the jet production efficiency is the largest at lowest values of 
$\lambda$, provided disk radiation efficiency is independent on $\lambda$.  

\noindent
$\bullet$
The lack of any signatures of correlation or anti-correlation of
radio source sizes with  Eddington ratio indicates the lack of any significant 
monotonic migration of objects (to lower or larger Eddington ratios);

\noindent
$\bullet$
A promising scenario which can explain energetics of jets in powerful radio 
sources and observed radio vs. optical luminosity correlations
is the one involving the magnetically arrested/choked
accretion flows. Such flows may support sufficiently large magnetic fluxes to
power jets with  $P_j \gtrsim L_{bol}$, while a truncation of accretion disks 
by a 'poloidal magnetosphere' can relax requirements of 
having $\eta_j \gtrsim 1$.

\noindent
$\bullet$
Our results suggest a connection of the cold accretion phase following a lower
accretion rate, hot accretion phase taking place in extragalactic radio 
sources.  
Such a two-phase scenario can overcome the difficulty of accumulating large 
magnetic fluxes by geometrically thin accretion disks. Without such a pre-phase,
the cold accretion events would not be accompanied by production of powerful 
jets.

\acknowledgments

MS and GM are grateful to R. Blandford and J. McKinney for many stimulating
discussions regarding magnetically-choked accretion scenarios.   
We acknowledge financial support by the Polish NCN grant DEC-2011/01/B/ST9/04845, 
by NASA Fermi grant no. NNX11AO39G, and by a Herschel Research Support 
Agreement (grant administered by NASA JPL) no. RSA 1433865.  GS and DKW acknowledge financial support from the European Associated Laboratory "Astrophysics Poland-France". NVA has been supported by CAPES (proc. no. 6382-10-0)

\clearpage

\vspace{5cm}

\begin{figure}
\centering
\includegraphics[width=0.95\textwidth]{P2_f1c--.ps}
\caption{Same as Fig. \ref{fig:fig1a} for the redshift range $0.2 < z <0.4$.}
\label{fig:fig1c}
\end{figure}

\clearpage

\begin{figure}
\centering
\includegraphics[width=0.95\textwidth]{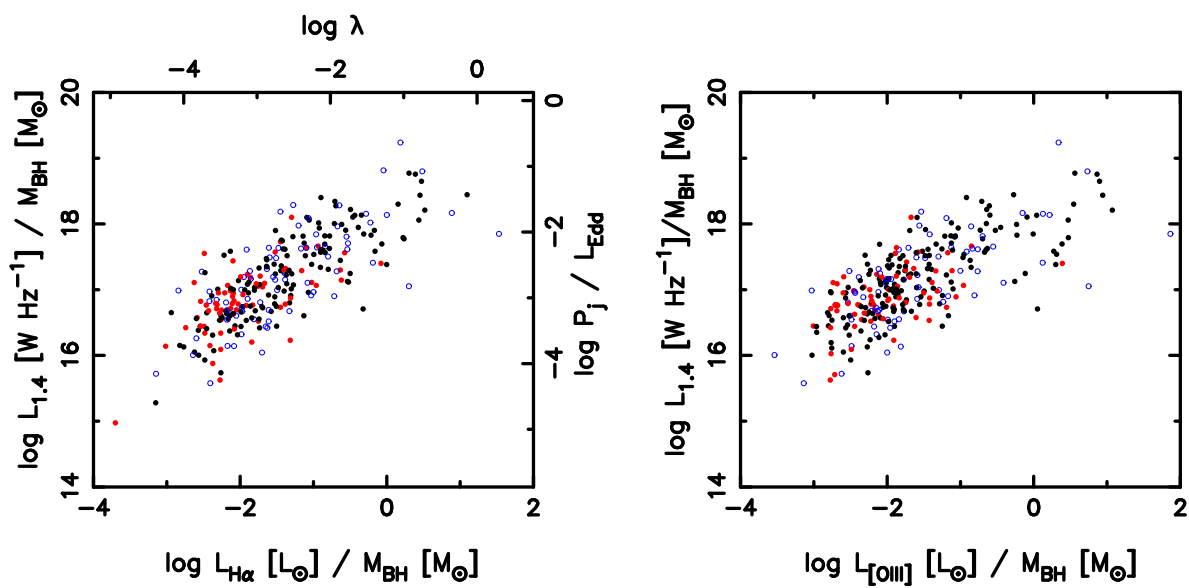}
\caption{The radio luminosities normalized by the black hole masses as a function of optical emission line luminosities (left panel:  $L_{\Ha}$, right panel: $L_{\oiii}$)
normalized by the black hole masses. The symbols have the same meaning as in Fig. 1. In the left panel are also indicated the values of the parameters log $\lambda$ and log $P_j/L_{edd}$ for $f=3$.}
\label{fig:fig2}
\end{figure}

\clearpage

\begin{figure}
\centering
\includegraphics[width=0.95\textwidth]{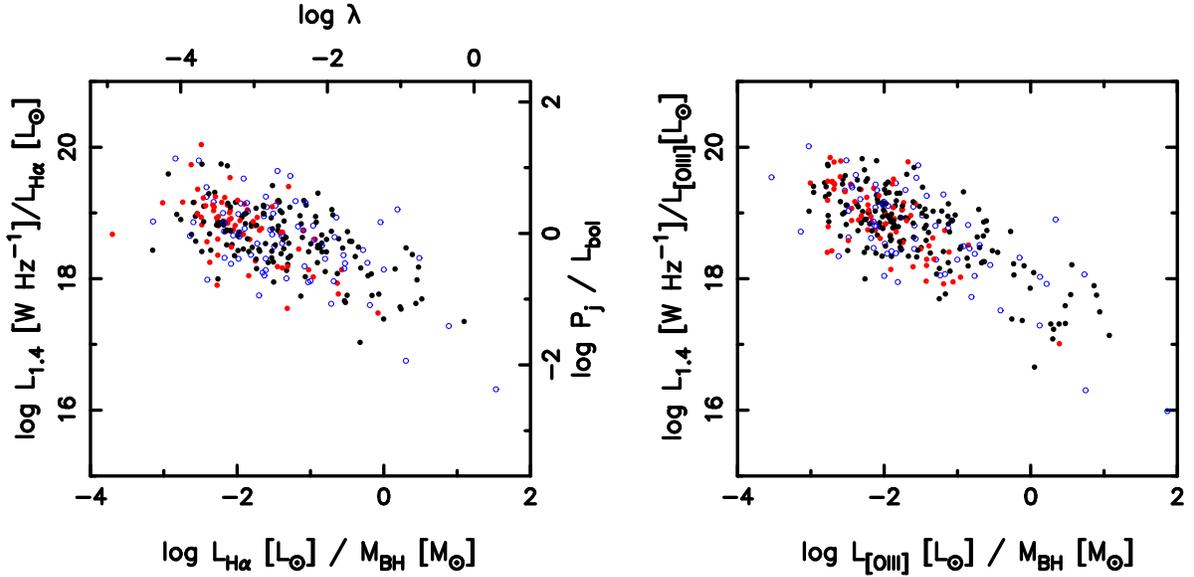}
\caption{ $L_{1.4}/L_{\Ha}$ as a function of  $L_{\Ha}/M_{BH}$ (left panel) and  $L_{1.4}/L_{\oiii}$ as a function of  $L_{\oiii}/M_{BH}$ (right panel). The symbols have the same meaning as in Fig. 1. In the left panel are also indicated the values of the parameters log $\lambda$ and log $P_j/L_{bol}$ for $f=3$.}
\label{fig:fig3}
\end{figure}

\clearpage

\begin{figure}
\centering
\includegraphics[width=0.95\textwidth]{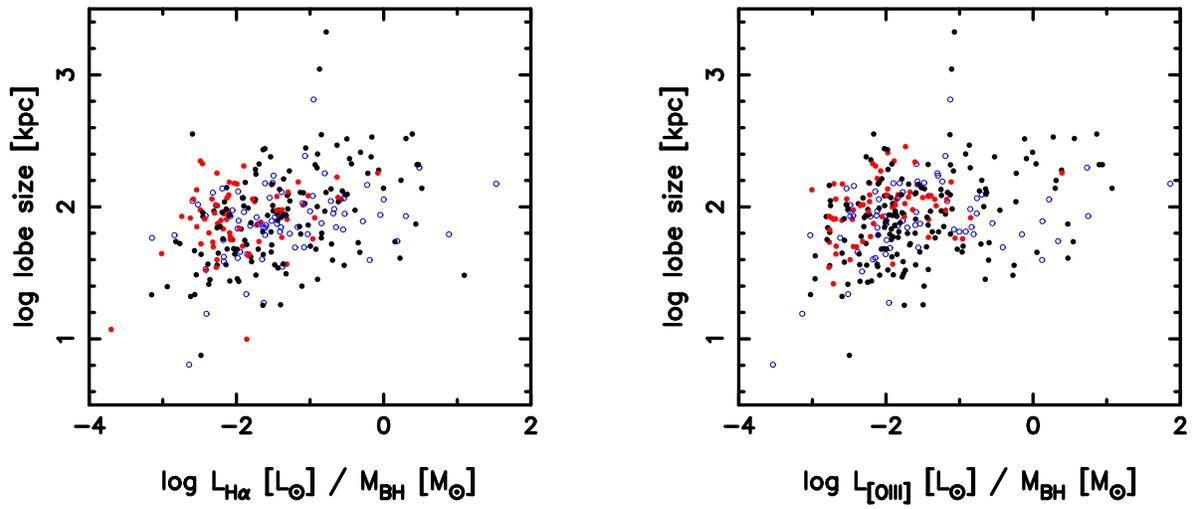}
\caption{The projected  sizes of the radio lobes as a function of the Eddington ratio, as measured using \Ha\ (left panel) and \oiii\ (right panel). The symbols have the same meaning as in Fig. 1.}
\label{fig:fig4}
\end{figure}

\begin{deluxetable}{cccccccccc}
\rotate
\tabletypesize{\scriptsize}

\tablewidth{550pt}

\tablenum{1}
\tablecaption{Radio and optical properties of the sample of radio galaxies\tablenotemark{1}}

\tablehead{\colhead{SDSS ID} & \colhead{Cambridge Cat. ID} & \colhead{Redshift} &\colhead{Radio Type} & \colhead{$log L_{1.4}$} & \colhead{$log L_{\Ha}$} & \colhead{$log L_{\oiii}$} & \colhead{$log _{M_{BH}}$} & \colhead{Ang. size} & \colhead{Lobe size} \\ 
\colhead{} & \colhead{}& \colhead{} & \colhead{} & \colhead{(W ${\rm Hz}^{-1}$)} & \colhead{($L_{\odot}$)} & \colhead{($L_{\odot}$)} & \colhead{($M_{\odot}$)} & \colhead{(arcsec)} & \colhead{(kpc)} } 

\startdata
0273.51957.633 & 4C +00.37 & 0.0968 & FRI & 25.65 & 6.839 & 6.644 & 8.87 & 170.64 & 151.02 \\
0312.51689.471 & 4C +00.56 & 0.0524 & FRII & 25.34 & 7.605 & 7.572 & 8.74 & 255.62 & 128.83 \\
0349.51699.169 & 6C B165818.4+630042 & 0.1063 & FRII & 25.45 & 6.417 & 6.579 & 7.83 & 139.2 & 135.45 \\
0366.52017.349 & 6C B171944.8+591634 & 0.2212 & FRII & 25.59 & 7.486 & 6.889 & 8.29 & 52 & 92.72 \\
0367.51997.294 & 4C +54.36 & 0.1852 & X-shaped & 26.00 & 7.195 & 6.740 & 8.16 & 78 & 121.12 \\
0385.51877.485 & 4C -00.83 & 0.1848 & FRI/II & 26.28 & 6.482 & 0.000 & 9.01 & 67.2 & 104.18 \\
0400.51820.424 & 4C +00.05 & 0.0793 & FRI/II & 25.37 & 6.127 & 0.000 & 8.55 & 45 & 33.71 \\
0432.51884.345 & 7C B073404.1+402639 & 0.3905 & FRII & 25.59 & 0.000 & 6.806 & 8.66 & 32 & 84.73 \\
0436.51883.010 & 6C B075738.1+435851 & 0.2554 & FRII & 25.66 & 6.899 & 6.740 & 8.42 & 26 & 51.63 \\
0439.51877.044 & 6C B080758.9+434635 & 0.1432 & X-shaped & 25.55 & 6.886 & 0.000 & 8.26 & 28 & 35.21 \\
0439.51877.436 & 7C B080310.1+452158 & 0.2439 & FRI & 25.06 & 0.000 & 7.105 & 8.17 & 30.46 & 57.97 \\
0439.51877.637 & 7C B081405.1+450809 & 0.1422 & FRII & 25.43 & 5.690 & 6.322 & 8.17 & 35 & 43.76 \\
0442.51882.241 & 6C B081421.2+500530 & 0.2804 & FRI/II & 25.93 & 7.272 & 0.000 & 8.67 & 33 & 70.12 \\
0442.51882.258 & 6C B081520.7+495611 & 0.0952 & One-sided & 24.87 & 6.227 & 6.426 & 8.20 & 45.03 & 78.53 \\
0448.51900.335 & 6C B084421.9+571115 & 0.1937 & FRII & 26.08 & 7.515 & 7.887 & 7.98 & 144 & 231.62 \\
0449.51900.323 & 7C B084921.8+544832 & 0.1133 & FRI & 25.57 & 6.519 & 0.000 & 8.38 & 9.78 & 9.95 \\
0450.51908.330 & 4C +56.17 & 0.1409 & FRII & 26.05 & 7.107 & 6.912 & 8.04 & 170 & 208.59 \\
0451.51908.541 & 7C B091959.0+571901 & 0.2846 & FRI & 25.24 & 6.973 & 6.500 & 8.72 & 51.55 & 109.78 \\
0484.51907.497 & 6C B090602.0+585910 & 0.2698 & X-shaped & 26.25 & 7.894 & 7.367 & 8.45 & 32.90 & 67.43 \\
0486.51910.456 & 7C B093527.5+622203 & 0.2298 & FRII & 25.41 & 0.000 & 6.277 & 8.65 & 58 & 105.59 \\
0487.51943.188 & 6C B095114.5+625546 & 0.2286 & FRI & 25.62 & 6.871 & 0.000 & 8.92 & 31.17 & 56.48 \\
0488.51914.191 & 6C B101400.2+634442 & 0.1839 & X-shaped & 25.72 & 7.021 & 7.431 & 8.43 & 42 & 64.85 \\
0490.51929.096 & 7C B105806.3+654923 & 0.1926 & FRII & 25.15 & 6.928 & 6.992 & 8.24 & 76.2 & 122.08 \\

\enddata

\tablenotetext{1}{Table 1 is published in its entirety in the electronic edition of ApJ. A portion is shown here for guidance regarding its form and content.}

\end{deluxetable}

\end{document}